**Phenomenological theory of clustering at the nuclear surface and symmetry energy**

Q. N. Usmani[a*], Zaliman Sauli[b] and Noraihan Abdullah[a]

[a]Institute of Engineering Mathematics, University Malaysia Perlis, Malaysia
[b]School of Microelectronics Engineering, University Malaysia Perlis, Malaysia

**Abstract**

We present a phenomenological theory of nuclei which incorporates clustering at the nuclear surface in a general form. The theory explains the recently extracted large symmetry energy at low densities of nuclear matter and is consistent with the static properties of nuclei. In phenomenological way clusters of all sizes, shapes along with medium modifications are included. The importance of quartic term in symmetry energy is demonstrated at and below the equilibrium density in nuclear matter. It is shown that it is related both to clustering as well as to the contribution of three-nucleon interaction to the equation of state of neutron matter. Reasons for these are discussed. Due to clustering the neutron skin thickness in nuclei reduces significantly.

[*] qnusmani@hotmail.com

Recently, large values of symmetry energy has been reported at nuclear matter densities $\rho \leq 0.009$ fm$^{-3}$ at low temperatures [1-2]. This arises because extra binding energies are gained due to cluster formation of various shapes and sizes in nuclear matter at sub-nuclear densities [3]. This finds explanation in Quantum Statistical (QS) [1, 4] approach which includes specific cluster correlations and then interpolates between the low density limit and the relativistic mean field (RMF) approaches near the saturation density. In this approach only clusters with A ≤ 4 has been included. In this letter, we introduce a thermodynamically consistent phenomenological approach in which cluster of all shapes and sizes along with medium modifications are included, *a priori*, by requiring that the binding energy of nuclear matter per nucleon in the limit of *average* zero density must approach its value at the equilibrium density. This is a conceptual requirement and was first pointed out in Ref. [5] in the study of $\alpha$-matter and then in [6] in connection with the virial expansion of the low density nuclear matter to demonstrate the $\alpha$-particle clustering. Obviously, such an approach will hide and/or not require a considerable amount of microscopic details such as in QS [1, 4] or other approaches [3, 7]. But the conclusions drawn based upon the present approach will be of general validity and guide us to explore new circumstances and regimes, e.g., to study isospin physics through hypernuclei [8, 9], besides the study of neutron stars and atomic nuclei through heavy-ion reactions [10]. For example, due to cluster formation, the Λ-binding to nuclear matter in the neighborhood of zero density must approach to its value at equilibrium density which is around 30 MeV – an outcome of the conceptual requirement mentioned above. This indeed is a fundamental departure from all the other earlier approaches [8, 11-13] and requires a separate study.

In addition to explaining the large values of symmetry energy at low densities in terms of clustering, a direct result of the present approach is that the slope of the symmetry energy in the neighborhood of zero density is negative, due to inclusion of clusters of all orders. On the other hand it is positive in the vicinity of equilibrium density. This implies that the symmetry energy should show at least one minimum in between the two limits, an interesting situation to be explored experimentally as described later. Another important outcome is the necessity of quartic term in isospin required for a consistent description of nuclei with a strength which is ≥ 15% of the quadratic term. It is demonstrated that in part it originates from the clustering at the nuclear surface, a fact also observed in the microscopic QS approach [4] and also due to contribution of three-nucleon interaction to the EoS of neutron matter for densities less than or equal to the equilibrium density. The theory must, of course, attend to other experimental data and theoretical constructs which are on firm footings. We realize this by including some of them as inputs to the theory.

We adopt an extended version of Thomas-Fermi (ETF) method which is based on the density functional approach [8, 14]. This technique has been extensively used in atomic, metallic clusters and nuclear physics and accurately produces the average part of the energy. Thus the quantal shell effects are smoothed out as in the liquid drop model or Strutinsky calculations. We obtain energies and charge *rms* radii of nuclei with a root mean square deviation ( $\sigma_E$ for energy and $\sigma_R$ for radii ) 0.937 MeV and 0.0237 fm, respectively, of their experimental values. Current microscopic-macroscopic and Hartree-Fock based theories [15, 16] give $\sigma_E \approx$ 0.6-0.7 MeV and $\sigma_R \approx$ 0.03 fm which include deformation and other phenomenological terms such as Wigner energy. Since we are addressing the question of large values of symmetry energies at low densities and significant quartic term in isospin, we consider our approach quite adequate to address these issues in meaningful ways. Our aim here is not to obtain a precise fit to masses, but rather take a pragmatic and swift path to see what we can infer and learn from clustering at low densities. Besides, the microscopic-macroscopic theories can not be generalized, not without major modifications, to include cluster formation at low densities. In our approach, large values of symmetry energies can be easily incorporated as described below.

Though, the present adopted version of ETF approach has been described earlier [8], but it is desirable that we give a few essential details for continuity and more importantly for the crucial required modifications to incorporate clustering at the nuclear surface. We write the energy of a nucleus as a functional of the density:

$$E[\rho] = \int \left[ \varepsilon(\rho) + \frac{\hbar^2}{72m}\left(\frac{\nabla\rho}{\rho}\right)^2 + \frac{\hbar^2}{6m}\nabla^2\rho + a_\rho \frac{(\nabla\rho)^2}{\rho} \right] \rho \, d\vec{r}$$
$$+ \int \left[ S(\rho)\delta^2 + Q(\rho)\delta^4 \right] \rho \, d\vec{r}$$
$$+ \frac{1}{2}e^2 \int \frac{\rho_p(\vec{r}_1)\rho_p(\vec{r}_2)}{|\vec{r}_1 - \vec{r}_2|} d\vec{r}_1 d\vec{r}_2 - \frac{3}{4}\left(\frac{3}{\pi}\right)^{1/3} e^2 \int \rho_p^{4/3}(\vec{r}) d\vec{r}$$
$$- Shell + a_{pair} \frac{(-1)^Z + (-1)^N}{A^{3/4}}$$

(1)

with $\delta = (\rho_n - \rho_p)/\rho$. The integral in the first line represents the volume and surface terms, the second integral is the contribution due to symmetry energy and the last two integrals are respectively the direct and exchange Coulomb energy. $S(\rho)$ and $Q(\rho)$ shall be described a little later. In the last line we have the quantal shell contribution which we extract from Ref [17]. The last term is the pairing energy contribution. Both, the shell and the pairing terms do not play significant roles, as far as the present study is concerned, but they improve the results quantitatively. $\rho_n$ and $\rho_p$ are respectively the neutron and proton densities, and $\rho$ is the total nucleon density; $\rho = \rho_n + \rho_p$. We have neglected the deformation effects as we consider only spherical or near spherical nuclei. $\varepsilon(\rho)$ is the equation of state of normal nuclear matter.

We consider, the following thermodynamically consistent picture of $\varepsilon(\rho)$. At the equilibrium density the nuclear matter (NM) is stable with $u_v$ MeV of binding energy per nucleon, where $u_v$ is the volume term in Bethe-Weisäcker mass formula. The lower densities of NM can be visualized through an isothermal expansion with rising energies per nucleon.

Still in the neighborhood of equilibrium density (for $\rho < \rho_0$), NM can be considered as a uniform medium, as is evident from the accurate auxiliary field diffusion Monte Carlo (AFDMC) calculations with Argonne $AV_6$' NN interaction [18]. $AV_6$' is a truncated version of $AV_8$' which is a simplified re-projection of the full $AV_{18}$. These calculations establish an important result. There is no sign of phase transition or formation of clusters for $\varepsilon(\rho)$ in the range $\rho = 0.08–0.16\ fm^{-3}$. Further isothermal expansion will eventually bring us to some density where the energy per nucleon will be a maximum and pressure zero, a region of unstable equilibrium. In the neighborhood of this region clusters begin to form. One can envision more cluster formation by lowering the density and energy of NM through further expansion; pressure is positive now. In this region, NM gives away its energy by performing external work. This process can be continued with the formation of larger and larger clusters and binding energies per nucleon, as revealed in QS and other approaches [1-7], till we reach the average zero density. In that limit $E/A$ again becomes $-u_v$ MeV and pressure zero. Our EoS for NM adheres to this picture.

We use quite general density functional for $\varepsilon(\rho)$ avoiding the use of specific NN or NNN interaction. Following Ref. [8], we write it as

$$\varepsilon(\rho_\geq) = -u_v + \frac{K}{18}\left(\frac{\rho - \rho_0}{\rho_0}\right)^2 + M\left(\frac{\rho - \rho_0}{\rho_0}\right)^3, \qquad \text{for } \rho \geq \rho_x \quad (2a)$$

$$\varepsilon(\rho_\leq) = -u_v + A\rho + B\rho^2 + C\rho^3 + D\rho^4 + \frac{3\hbar^2 (3\pi^2)^{2/3}}{5m_N}\rho^{2/3}. \quad \text{For } \rho \leq \rho_x \quad (2b)$$

Notice in 2b, when the density approaches zero, the binding energy per nucleon becomes $u_v$. Our $\varepsilon(\rho)$ follows the general pattern as a function of density as explained in the previous paragraph.. The constant terms $A$, $B$, $C$ and $D$ are determined by equalizing $\varepsilon(\rho_\geq)$ and $\varepsilon(\rho_\leq)$ and their derivatives at $\rho = \rho_x$, where $\rho_x$ is a density parameter between 0 and the equilibrium nuclear matter density $\rho_0$. One may use a single expression in the entire density range. Instead of using 2b, one may add more terms in the Taylor series expansion of 2a and then put the condition that at $\rho = 0$, $K/18 + M + \cdots = 0$. This will ensure that the binding energy per nucleon of NM approaches $u_v$ for $\rho = 0$. There can be other alternatives also, for example, use of Skyrme-type density functionals but with the proviso $\varepsilon(\rho = 0) = -u_v$ by adding phenomenological terms. However, we find convenient to work with (2a) and (2b) as it provides a simple control on EoS. It should be emphasized that (2) contains mainly two parameters $M$ (related to asymmetry in the saturation curve of NM) and $\rho_x$. The other parameters are consistent with the generally accepted values, $u_v$=16 MeV, the compression modulus $K = 250$ MeV and $\rho_0 = 0.16\, fm^{-3}$, though we do vary these also. For the neutron and proton densities, we use two parameter Fermi distribution for each species [8].

To be consistent with [1], we define the symmetry energy as the difference between energy per nucleon of neutron matter and the nuclear matter as given by (2). Since neutron matter is a superfluid gas with positive pressures at all densities, it immediately gives $S(\rho \to 0) + Q(\rho \to 0) \to u_v$. This definition of symmetry energy differs from the standard alternative where the symmetry energy is defined as the second derivative of the energy density with respect to isospin asymmetry $\delta$ [10, 19], which is related to experimental observables for nuclei near to $N = Z$. The two definitions become identical if $Q(\rho) = 0$. At present, we have no idea regarding the density dependence of the quartic term in the symmetry energy. We thus assume the same dependence as for the quadratic term and replace $S(\rho)\delta^2 + Q(\rho)\delta^4$ by $(1-q)E_{sym}(\rho)\delta^2 + qE_{sym}(\rho)\delta^4$ in (1). The parameter $q$ determines the relative importance of the two terms. This parameter is found to play an important role in giving good fit to the binding energies and root mean square radii (*rms*) for our severely constrained EoS of NM and symmetry energy. It turns out that if we put $q = 0$ then we have to turn off clustering at the nuclear surface as well as the three-nucleon interaction in the EoS of neutron matter to account for the static properties in a reasonable way as demonstrated later.

For the neutron matter EoS, we employ the recently calculated values [20]. This has been obtained by employing an accurate fixed phase AFDMC technique with 66 neutrons enclosed in a periodic box with Argonne $AV_8$' [21] and Urbana three-nucleon UIX [22] interactions. There is little difference between the results of neutron matter for $AV_8$' and $AV_{18}$' in the low density region. Similarly, in this region, it may also expected that results due to use of Urbana UIX three-body interaction may be close to the more sophisticated Illinois IL2 interaction [23] which is required to produce the ground and excited state energies of p-shell nuclei in the GFMC calculations [24]. However, this needs confirmation. In Fig. 1 (left panel) we plot the results of Ref. [20], represented by filled circles for $AV_8$'+ UIX. The solid line is the fit obtained by $E(\rho)/A = \sum_{i=1,3} y_i \rho^i / \left(1 + \sum_{i=1,4} z_i \rho^i \right)$, where the parameter values are $y_1$=0.330501 x $10^4$, $y_2$= 0.631598 x $10^7$, $y_3$=0.259058 x $10^9$, $z_1$= 0.947918 x $10^4$  $z_2$= 0.192323 x $10^7$,  $z_3$=0.772199 x $10^7$, $z_4$= –0.323126 x $10^8$. The open circles represent the results with $AV_8$' alone and can be obtained by multiplying the solid curve with a fudge factor $exp(-2.615(\rho-0.05))$ for $\rho > 0.05$ $fm^{-3}$. We use these fits in our calculations of $E_{sym}(\rho)$.

We have a total of eight parameters, namely, $K$, $u_v$, $\rho_0$, $M$ and $\rho_x$ in (2), $a_\rho$ which controls the surface, $q$ and $a_{pair}$ in (1). We vary seven of them at a time for specific values of $\rho_x$ through a standard minimization procedure to produce the experimental *rms* radii [25] and energies [26]. Calculated energies are obtained variationally by varying the density. Since our aim is to study the physics related to clustering (and not to fit large number of masses in the entire mass table), we confine to a total of 376 spherical nuclei [15, 26-27] from $^{12}C$ to $^{219}U$. They include the chains $^{38-52}Ca$, $^{42-54}Ti$, 100-134$Sn$ and $^{178-214}Pb$. For the charge rms radii we have considered 50 nuclei. In Table 1, we compare our fits with various other approaches. Column 3 gives the result from Ref. [28]. In certain respects, this approach is similar to ours but without incorporating clustering. Column 4, gives the results in the Skyrme-Hartree-Fock-Bogoliubov microscopic-macroscopic approach [16] but also without clustering. The last two columns, give results for liquid drop models and their various versions with quantal corrections and deformation. Considering that we have not included Wigner energy contribution and sophisticated pairing energy terms, as in Refs. [15, 16, 27], our approach works very well. Our *rms* radii are better than those in the other approaches, though our binding energies are not that good. In Fig. 2, we plot the differences between the calculated (*cal*) and experimental (*exp*) energies (left panel) and the proton *rms* radii (right panel). These values are plotted for our preferred $\rho_x = 0.06$ *fm*. We varied $\rho_x$ between 0.025 and 0.12 *fm*$^{-3}$. The binding energies and *rms* radii are not very sensitive to $\rho_x$, but the symmetry energies are in the low density region. In the right panel of Fig. 1, we plot the EoS for NM for $\rho_x = 0.05$ *fm*$^{-3}$ (green, short-dashed curve), $\rho_x = 0.06$ *fm*$^{-3}$ (red, solid curve) and $\rho_x = 0.07$ *fm*$^{-3}$ (blue, long-dashed curve). For a change of $\rho_x$ by 0.02 *fm*$^{-3}$ in this range, the change in the location of maximum in the EoS of NM is only 0.005 *fm*$^{-3}$, which is the region of unstable equilibrium. Thus it is pretty much fixed around $\rho = 0.026$ *fm*$^{-3}$ and indicates the onset of clustering around and below this density.

In Fig. 3, the results for the symmetry energies are given. The color code and legends for the various curves are same as those in Fig. 1 (right panel) described earlier. The dotted curve depicts the results of QS approach [1, 29] at T = 1 MeV. The experimental extraction of the symmetry energy was obtained in Ref. [1, 2], in the low density region in the pioneering experiment on heavy ion collisions of $^{64}$Zn on $^{92}$Mo and $^{197}$Au at 35 MeV/A. The down blue triangles are the data from [1] which were obtained after correcting it for energy recalibration and reevaluation for particle yields in different velocity bins. They are therefore slightly different from [2]. We have shown an error bar of ±15% as reported in [2].

Symmetry energy is not a directly measurable quantity. It is extracted indirectly from other observables which depend on the symmetry energy, thus some model dependence or dependence on theoretical interpretations is inevitable. Significantly, the medium effects on the clusters play an important role [30, 31]. The up red triangle, are the data from [1] which were corrected for the medium effects in a self consistence way. The whole bunch of data points (down blue triangles) shifts to considerably higher densities (up red triangles) and there is an upward trend for the symmetry energies for lower densities, Fig. 3; the down blue triangles have a downward trend. The slope of our calculated curves, represented by short-dashed, solid and long-dashed lines are all negative at low densities as a result of our ansatz (2) and the EoS of neutron matter. This is in conformity with the data; the up red triangles which have been corrected for medium effects. Clearly, our calculations distinguish between the two sets of data (the up red and down blue triangles). Our symmetry energy shows a distinct minimum at $\rho_{min} \approx 0.02\, fm^{-3}$. Above this density the quasi-particle picture dominates and below this density the cluster formation takes over. In QS approach, this minimum is not seen, simply because heavier clusters are not included. Thus, it is important that this region of density should be explored experimentally. The right panel of Fig 3 also give the experimental symmetry energy data as obtained from studies in heavy ion multi-fragmentation reactions at relatively higher densities. The red and green circles are from [32], the blue circle from [33] and the pink square is from [34] at the equilibrium density. The calculated curves all lie somewhat above the experimental values including the QS-RMF [29]. A possible reason for this discrepancy could be traced to the quartic term. The experimental extraction of the symmetry energy assumes only a quadratic dependence.

Presence of a quartic term in symmetry energy at high densities has been proposed earlier [19] which strongly modifies critical density for the direct Urca process in connection with the cooling of neutron stars. Its presence in the low density region is attributable to clustering at the nuclear surface and the contribution of three-nucleon interaction near the equilibrium density in the EoS of neutron matter. To demonstrate this we give in table II, fits for various situations pertaining to clustering (Yes, $u_v \approx 16$ MeV in 2b) and no-clustering (No, $u_v \approx 0$ MeV in 2b) as well as with and without UIX. Columns 3 and 4 give respectively the root mean square deviations $\sigma_E$ and $\sigma_R$ (*rms*). In the first row of results, where *q* was put equal to zero, *i e*. no quartic term in the isospin, the $\sigma$ values are very large. Varying *q*, second row, gives a dramatic reduction by a factor of $\approx 7$ for $\sigma_E$ and a factor of $\approx 4$ for $\sigma_R$. This amply justifies the inclusion of quartic term and signifies its importance. It is also evident from the results given in the next two rows that both clustering and the three-nucleon interaction in the EoS of neutron matter are responsible for the appearance of the quartic term. The last row roughly mimics the mean field calculations (Skyrme-Hartee-Fock (Sky-HF) and RMF). For this and the mean field theories the symmetry energy goes to zero as $\rho \to 0$. Here, there is no clustering, no three-nucleon interaction in EoS of neutron matter and almost no quartic term. In a nucleus near $A \approx 100$, the binding energy is around 850 MeV. A deviation $\sigma_E$ by 1 MeV from the experiment amounts to $\approx 0.1$ %. On the other hand, in neutron matter at $\rho_0$ [Fig. 1, left panel] the contribution of three-nucleon interaction is $\approx 20$ % of *E/A*. Similarly, in p-shell nuclei, its contribution is 15-20% for $^6Li$–$^{12}C$ going up with increasing mass number [24,35]. It contributes grossly to energies of nuclei and neutron matter. Therefore a $\sigma_E$ of around 1 MeV ($\approx 0.1$ %) provides a good systematic to study the effects of three-nucleon force. To the best of our knowledge, in all the earlier calculations, the role of three-nucleon interaction has been either totally ignored or not treated properly, otherwise the quartic term in isospin would have been seen earlier. The static property of nuclei are insensitive to quartic term even in the precision fittings of microscopic-macroscopic models; it appears only because of the constrain or use of accurate EoS of neutron matter and the incorporation of clustering that we have modeled through (2).

Though, some of the microscopic-macroscopic models do contain non-quadratic term in isospin in the form of Wigner energy, but they have an entirely different origin. The origin of quartic term due to potential energies at high densities was emphasized in Ref. [19] which we witness here near or below equilibrium densities due to three-nucleon interaction as evident from the second and third row of results of table II. The origin of the quartic term due to clustering is perhaps not surprising. The kinetic energy in the symmetry energy are known to have quartic parts, which gets enhanced due to clustering in the low density region. As is also evident from table II, results with clustering are significantly better than with no-clustering.

A quantity of interest is the neutron skin thickness [36], defined as the difference between the *rms* radii of neutrons and protons $\delta R = \sqrt{\langle r_n^2 \rangle} - \sqrt{\langle r_p^2 \rangle}$. In Sky-HF theories $\delta R$ is sensitive to the slope of the symmetry energy, $L$ at the equilibrium density, which is defined as $L = 3\rho_0 \left. \frac{\partial S_{sym}}{\partial \rho} \right|_{\rho_0}$. We expect the clustering to affect $\delta R$ significantly as it is a direct surface phenomenon. In Table III, we give results of our calculations for clustering and no-clustering. Results for no-clustering are in reasonable agreement with Sky-HF [10, 37] and RMF[38] calculations, and experimental deductions [39-42]. However, there is clear discrepancy between the clustering and the other results including experiment. We find much lower values for $\delta R$. Does this imply that experimental deductions are implemented assuming no-clustering? They are indeed model dependent. The proposed parity-violating electron scattering experiment [43,44] at the Jefferson Laboratory will greatly help to clarify this. Our $L$ value for both the cases of clustering and no-clustering is same; $L \approx 68$ MeV which is well within the range of values extracted from isospin diffusion data. The error bars on $q$ in table II and $\delta R$ in table III reflects mainly the uncertainties on $\rho_x$ which is ±0.01. For $\rho_x$=0.06±0.01, other values of parameters are: $q = 0.16±0.01$, $u_v$ =16.00 MeV, $\rho_0 = 0.16$ *fm*⁻

[3], $K = 251.55$ MeV, $M = -8.71 \pm 1.50$ MeV, $a_\rho = 45.14 \pm 0.03$ MeV and $a_{pair} = 36.1 \pm 1.0$ MeV. Our value of $K$ is slightly higher that the values extracted from the Isoscalar Giant Monopole Resonances [44] $\approx 230$ MeV, but as argued in Ref. [15], this is not significant as it changes the $\sigma$ by 0.5%.

Clearly, the structure of $\varepsilon(\rho)$ for low densities is quite intricate which we have modeled through (2). The constant terms $A$, $B$, $C$ and $D$ in (2b) can be expressed in terms of $K$, $M$, $\rho_0$ and $\rho_x$ by equalizing $\varepsilon(\rho)$ and its first three derivatives at $\rho_x$. This ensures that $K$ and its first derivative is continuous for $\rho \leq \rho_x$. This may not be so, for instance the compression modulus $K$ may not be even continuous at some low density. But this can only be decided through an accurate many-body calculation employing, for example, the AFDMC technique for symmetric NM at low densities, or advancement in QS and other approaches [1-7] may shed some light. We have to wait till such calculations become feasible or clear experimental signatures of phase transition at low densities are established for $N \approx Z$.

In conclusion, we have presented a unified theory of nuclei which is reasonably consistent with the static properties of nuclei as well as clustering at the nuclear surface and incorporates the large values of the symmetry energies at low densities. Two main conclusions are: (a) The slope of the symmetry energy is negative at low densities and (b) establishes that quartic term in isospin plays a very important role; it originates from clustering as well as due to three-nucleon interaction. In addition, we have also demonstrated that cluster formation begins for $\rho$ around $0.026 \, fm^{-3}$ and the symmetry energy has a minimum at $\rho \approx 0.02 \, fm^{-3}$ below which clustering starts dominating.

QNU acknowledges useful correspondence with S. Shlomo, G. Röpke, K. E. Schmidt and A. W. Steiner.

References:

1. J. B. Natowitz, *et al*, Phys.Rev.Lett. 104, 202501 (2010).
2. S. Kowalski *et al.*, Phys. Rev. C **75**, 014601 (2007).
3. G. Watanabe *et al.*, Phys. Rev. Lett. **103**, 121101 (2009).
4. S. Typel, G. Röpke, T. Klähn, D. Blaschke and H. H. Walter, Phys. Rev. C **81**, 015803 (2010)
5. M. T. Johnson and J. W. Clark, Kinam **2** 3(1980).
6. C. J. Horowitz and A. Schwenk, Nucl. Phys. A **776**, 55 (2006).
7. H. Shen, H. Toki, K. Oyamatsu and K. Sumiyoshi, Nucl. Phys. A **637**, 435 (1998).
8. Q. N. Usmani, A. R. Bodmer and Z. Sauli, C **77**, 034312 (2008).
9. E. Hiyama, Y. Yamamoto, T. Motoba and M. Kamimura, Phys. Rev. C **80**, 054321 (2009).
10. B. A. Li, L. W. Chen and C. M. Ko, Phys. Rep. **464**, 113 (2008).
11. Q. N. Usmani and A. R. Bodmer; Phys. Rev. C**60**, 055215 (1999) .
12. D. J. Millener, C.B.Dover and A.Gal, *Phys. Rev.* **C38**, 2700 (1988).
13. O. Hashimoto and H. Tamura, Prog. In Part. and Nucl. Phys. **57**, 564 (2006) and references therein.
14. M. Centelles, P. Schuck and X. Vinas, Ann. Phys. **322**, 363 (2007).
15. P. Möller, J. R. Nix and K. L. Kratz, At. Data Nucl. Data Tables **66**, 131 (1997). P. Möller, J. R. Nix, W. D. Myers and W. J. Swiatecki, At. Data Nucl. Data Tables 59, 185 (1995).
16. S. Goriely, N. Chamel, and J. M. Pearson, *Phys. Rev. Lett.* **102**, 152503 (2009) and references their in.
17. W.D. Myers and W.J. Swiatecki, LBL Report 36803, 1994; W.D. Myers, W.J. Swiatecki, Nucl. Phys., **A 601**, 141 (1996).
18. Stefano Gandolfi, Francesco Pederiva, Stefano Fantoni, and Kevin E. Schmidt, Phys. Rev. Lett. **98,** 102503 (2007).
19. Andrew W. Steiner, Phys. Rev. C **74** 045808 (2008).
20. S. Gandolfi, A. Yu. Illarionov, K. E. Schmidt, F. Pederiva and S. Fantoni, Phys. Rev. C **79** 054005 (2009).
21. R. B. Wiringa, V. G. J. Stoks and R. Schiavilla, Phys. Rev. **C51** 38 (1995).

22. B. S. Pudliner, V. R. Pandharipande, J. Carlson and R. B. Wiringa, Phys. Rev. Lett. **74** 4396 (1995).
23. S. C. Pieper *et al.*, Phys. Rev. C **64** 014001 (2001).
24. S. C. Pieper and R. B. Wiringa, Annu. Rev. Part. Sci. **51** 53 (2001).
25. I. Angeli, Atomic Data and Nuclear Data Tables **87** 185 (2004).
26. G. Audi, A. H. Wapstra and C. Thibault, Nucl. Phys. **A729** 337 (2003).
27. A. Bhagwat, X. Viňas, M. Centelles, P. Schuck and R. Wyss, Phys. Rev. C **81** 044321 (2010).
28. M. Baldo, P. Schuck and X. Viňas, Phys. Lett. B **663** 390 (2008).
29. G. Röpke, private communication.
30. G. Röpke, Phys. Rev. C **79** 014002 (2009).
31. S. Shlomo *et al.*, Phys. Rev. C **79** 034604 (2009).
32. D. V. Shetty, S.J. Yennello and G.A. Souliotis, Phys. Rev. C **76** 024606 (2007).; D. V. Shetty *et al.*, J. Phys. G **36** 075103 (2009).
33. L. Trippa, G. Colo and E. Vigezzi, Phys. Rev. C **77** 061304 (2008).
34. D. T. Khoa and H.S. Than, Phys. Rev. C **71** 044601 (2005); D. T. Khoa, H.S. Than and D.C. Cuong, Phys. Rev. C **76** 014603 (2007).
35. Steven C Pieper, K. Varga and R. B. Wiringa, Phys. Rev. C **66** 044310 (2002); Steven C. Pieper, Nucl. Phys. A **751** 516 (2005).
36. A. R. Bodmer and Q. N. Usmani, Phys. Rev. C **67** 034305 (2003).
37. L. W. Chen, C. M. Ko, B. A. Li, Phys. Rev. C **72** 064309 (2005); A. W. Steiner and B. A. Li, Phys. Rev. C **72** 041601 (R) (2005), B. A. Li, A. W. Steiner, Phys. Lett. B **436**, 436 (2006).
38. K. Yako, H. Sagawa, H. Sakai, Phys Rev. C **74** 051303(R) (2006).
39. J. Piekarewicz, Phys. Rev. C **69** 041301 (2004); B.G. Todd-Rutel, J. Piekarewicz,, Phys. Rev. Lett. **95** 122501 (2005).
40. B. Klos, et al., Phys. Rev. C **76** 014311 (2007).
41. A. Klimkiewicz, et al. (LAND Collaboration), Phys. Rev. C **76** 051603(R) (2007).
42. S. Terashima, et al., Phys. Rev. C **77** 024317 (2008).
43. C.J. Horowitz, J. Piekarewicz, Phys Rev. C **63** 025501 (2001).
44. Jefferson Laboratory Experiment E-00-003, spokesperson R. Michaels, P.A. Souder, and G.M. Urciuoli.
45. J.P. Blaizot, Nucl. Phys. A **649,** 61c (1999).

| rms deviation | Present | Ref. [28] | HFB-17 [16] | LDM+WK[27] | Ref. [15] |
|---|---|---|---|---|---|
| $\sigma_E$ MeV | 0.937 | 1.7 | 0.581 | 0.630 | 0.669 |
| $\sigma_R$ MeV | 0.023 | 0.031 | 0.030 | – | – |
| No. of Nuclei | 376 | 161 | 2149 | 367 | 1654 |

Table I: Root mean square deviations in various approaches.

| Neutron Matter | Clustering | $\sigma_E$ MeV | $\sigma_R$ fm | $q$ |
|---|---|---|---|---|
| AV8'+UIX | Yes | 7.25 | 0.080 | 0.000 |
| AV8'+UIX | Yes | 0.936 | 0.023 | 0.160±0.004 |
| AV8'+UIX | No | 1.368 | 0.022 | 0.099±0.006 |
| AV8' | Yes | 0.902 | 0.023 | 0.076±0.005 |
| AV8' | No | 1.290 | 0.023 | 0.011±0.005 |

Table II: See text for details

| Nucleus | $\delta R$ (fm) | | | | |
|---|---|---|---|---|---|
| | Clustering | No-clustering | Sky-HF [10,37] | RMF [38] | Exp. |
| $^{208}Pb$ | 0.102±0.025 | 0.153±0.030 | 0.22±0.04 | 0.21 | 0.16±0.06 [40] |
| | | | 0.15 [16] | | 0.18±0.035 [41] |
| $^{132}Sn$ | 0.157±0.025 | 0.225±0.028 | 0.29±0.04 | 0.27 | 0.24±0.04 [41] |
| $^{124}Sn$ | 0.107±0.025 | 0.163±0.025 | 0.22±0.04 | 0.19 | 0.185±0.017 [42] |
| $^{90}Zr$ | 0.041±0.014 | 0.075±0.015 | 0.088±0.04 | – | 0.07±0.04 [39] |
| $^{48}Ca$ | 0.105±0.020 | 0.171±0.018 | – | – | – |

Table III: All entries are in Fermis.

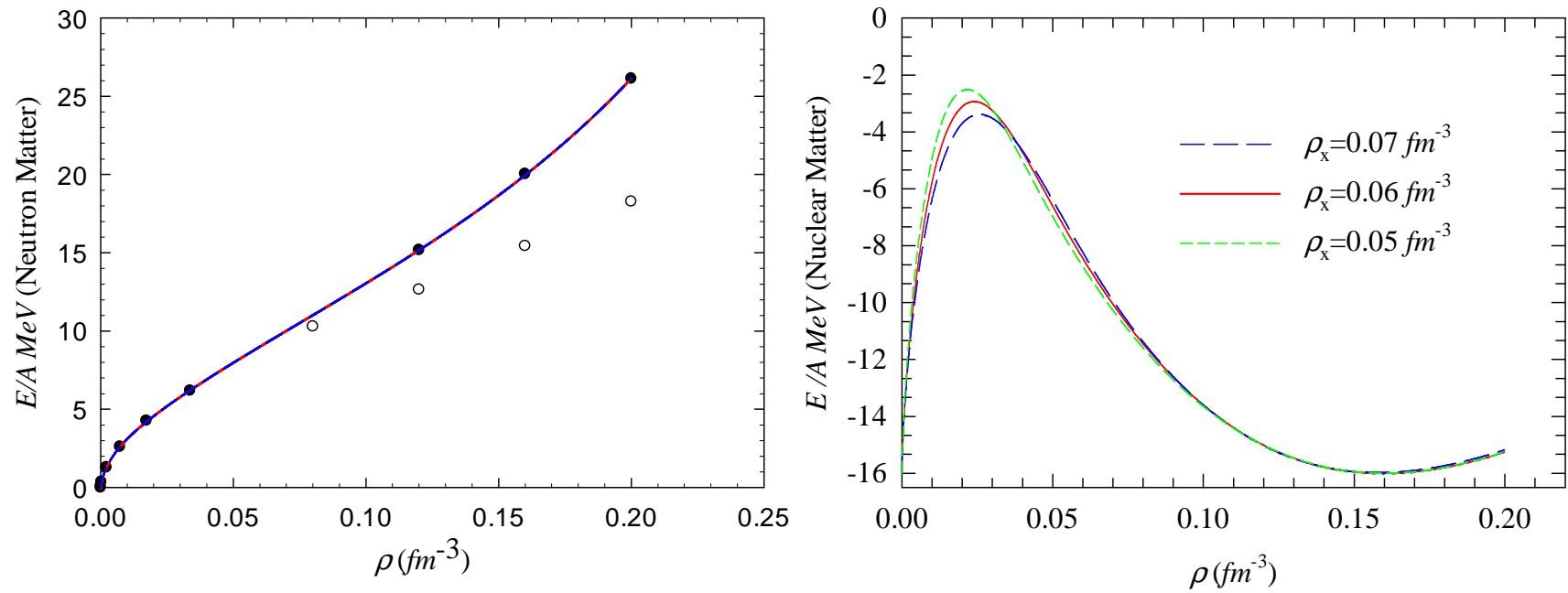

Fig.1 (Color online): Equation of State of neutron and nuclear matter as function of density.

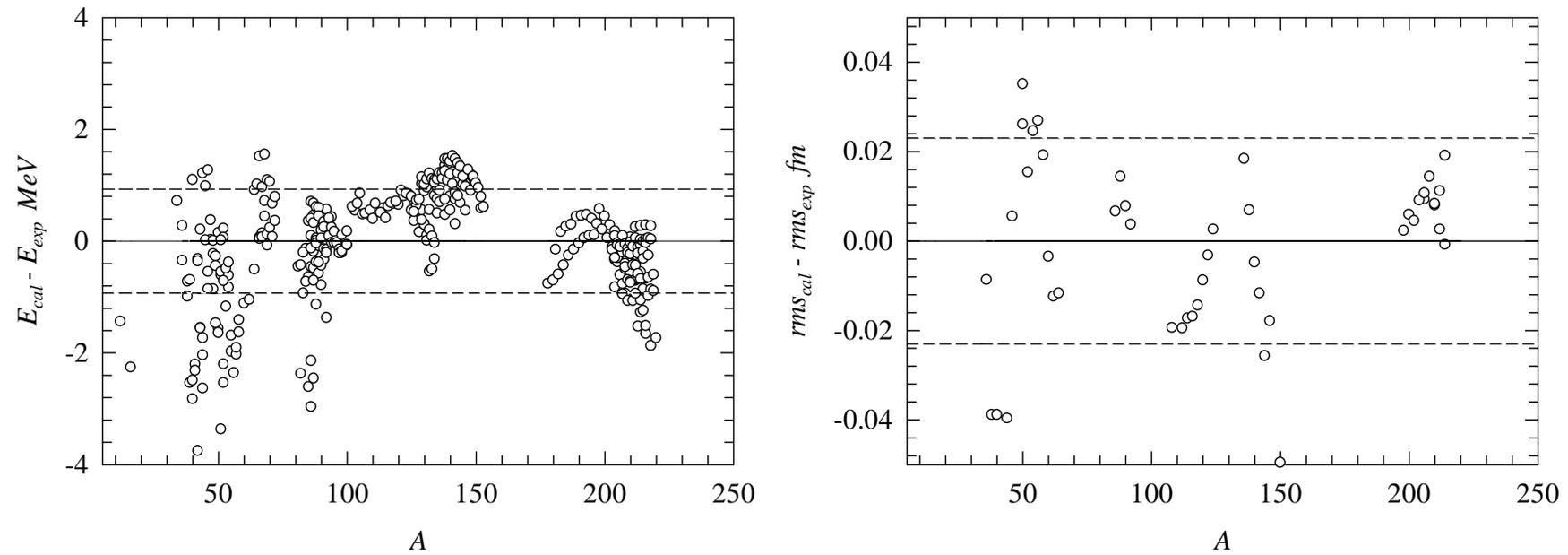

Fig. 2: Calculated and experimental differences for energies and proton charge *rms* radii as a function of *A*.

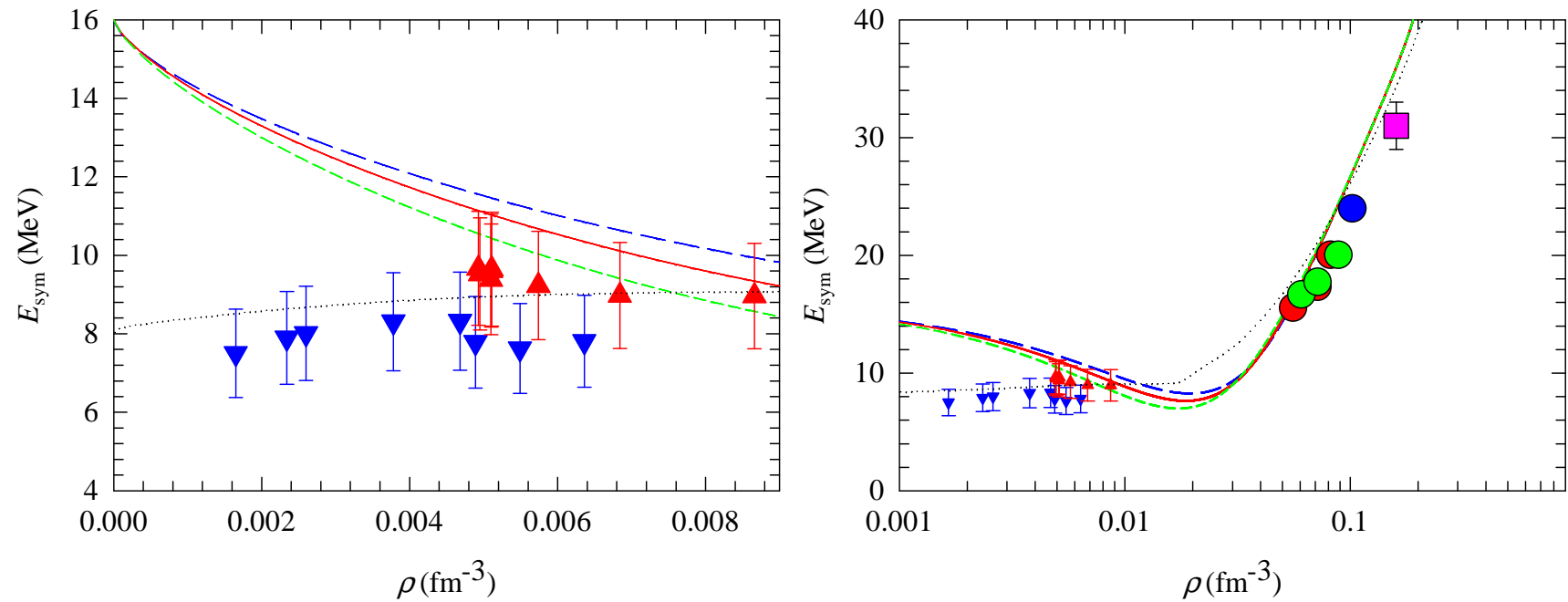

Fig. 3 (Color online): Symmetry energy as function of density. For details, see text.